\documentclass[%
 reprint,
 amsmath,amssymb,
 aps,
 floatfix,
]{revtex4-2}

\usepackage{graphicx}
\usepackage{bm}
\usepackage{hyperref}

\begin{document}

\title{Evolutionary Le Chatelier's Principle: Phenotypic Plasticity and Genetic Assimilation via Timescale Separation in the Price Equation}

\author{Tetsuhiro S. Hatakeyama}
\email{hatakeyama@elsi.jp}
\affiliation{Earth-Life Science Institute, Institute of Future Science, Institute of Science Tokyo, 2-12-1-IE-1 Ookayama, Meguro-ku, Tokyo 152-8550, Japan}

\date{July 2026}

\begin{abstract}
Phenotypic plasticity and genetic assimilation play key roles in adaptive evolution, yet their underlying mechanism has lacked a unified physical description.
A major theoretical difficulty lies in the fundamental difference in timescales, as phenotypic plasticity occurs rapidly within a generation whereas genetic changes accumulate slowly across generations. 
Here, we formalize these processes by bridging the continuous-time Price equation, a foundational equation of evolutionary dynamics, with the physical concept of timescale separation. 
A sudden environmental change induces a fast plastic displacement of the phenotype relative to the slow genotypic variable.
Through genotype--phenotype coupling, this displacement generates an internal genetic stress. 
We demonstrate that genetic assimilation is a dynamical relaxation process in which the genotype evolves to resolve this self-generated stress. 
These evolutionary dynamics mathematically realize Le Chatelier's principle, where the slow genetic response naturally amplifies the initial plastic shift in the same direction. 
The theory predicts that a weaker restoring force, which can manifest as larger clonal phenotypic fluctuations, requires a longer evolutionary timescale for assimilation. 
In the ideal limit of cost-free, perfectly adaptive plasticity, the relaxation time diverges, so assimilation effectively stalls. 
This formulation provides a macroscopic physical mechanism for genetic assimilation, offering a universal response law for evolutionary systems in which rapid phenotypic responses precede slower genetic change.
\end{abstract}

\maketitle

\section{Introduction}

Phenotypic plasticity allows organisms to rapidly alter their traits in response to sudden environmental changes, providing a buffer for survival \cite{pigliucci2001phenotypic}. 
From a biophysical perspective, plasticity can be viewed as a strategy to process environmental information and optimize population growth in fluctuating environments \cite{kussell2005phenotypic, rivoire2011value}. 
Its evolutionary role extends beyond survival; it is increasingly recognized that plasticity can act as a driver of adaptive evolution \cite{west2003developmental, levis2016evaluating}. 
A classic manifestation of this is genetic assimilation, first proposed by Waddington \cite{waddington1942canalization, waddington1953genetic, waddington1957strategy} and Schmalhausen \cite{schmalhausen1949factors}. 
In this process, a phenotypic trait initially expressed via plasticity in response to an environmental shift often represents an incomplete adaptation to novel conditions \cite{ghalambor2007adaptive}. 
This trait is later genetically stabilized, eventually becoming constitutive even in the absence of the original environmental trigger \cite{waddington1942canalization, crispo2007baldwin} or direct selection for the phenotype itself \cite{masel2004genetic}. 
While this phenomenon has been empirically demonstrated \cite{waddington1953genetic, suzuki2006evolution}, the mechanistic link between the initial plastic response and the subsequent genetic evolution remains a subject of ongoing debate. 
Furthermore, while maintaining and expressing phenotypic plasticity often incurs inherent fitness costs \cite{dewitt1998costs, murren2015constraints}, the dynamic role of this plasticity cost in driving genetic assimilation has not been fully formulated.

The central difficulty in understanding genetic assimilation as a general phenomenon lies in the difference in timescales between phenotypic and genetic processes. 
Phenotypic plasticity is a rapid, intra-generational response, while genetic changes accumulate slowly across generations; the dynamic interplay between these separated timescales has long been recognized as a key driver of evolution \cite{slobodkin1974optimal}. 
Classical quantitative genetics \cite{lande1979quantitative, lande2009adaptation, chevin2010adaptation} and recent mechanistic models \cite{ancel2000plasticity, espinosa2011phenotypic, raju2023theoretical} describe these evolutionary trajectories by assuming specific microscopic genetic architectures or developmental processes. 
For instance, quantitative models often treat the degree of plasticity as an evolving genetic locus, and mechanistic models rely on explicit equations for developmental networks to simulate evolutionary updates. 
These approaches depend on microscopic details of the system, which makes it difficult to identify the underlying physical dynamics of assimilation. 
While Waddington intuitively conceptualized such macroscopic dynamics through the metaphor of an ``epigenetic landscape'' \cite{waddington1957strategy}, a macroscopic formulation based on timescale separation is required. 
This framework should describe the response to external perturbations independently of microscopic details.

In physics, thermodynamics provides macroscopic laws that inherently incorporate timescale separation. 
Specifically, the Le Chatelier-Braun principle (hereafter referred to simply as Le Chatelier's principle) dictates that a system's long-term response to an external perturbation is amplified when its internal slow variables are allowed to relax, compared to its short-term response where these variables are constrained \cite{callen1985thermodynamics}. 
This principle is a mathematical property of any system undergoing constrained optimization, including economic systems \cite{samuelson1947foundations}. 
Given the mathematical analogies between evolutionary dynamics and statistical mechanics \cite{iwasa1988free, sella2005application, kobayashi2015fluctuation}, applying Le Chatelier's principle to evolutionary responses has been a viable hypothesis. 
Previous works have argued for the necessity of macroscopic theories for evolving systems \cite{goldenfeld2011life}, and indeed a static version of the evolutionary Le Chatelier's principle has been suggested based on static potential \cite{furusawa2015global, kaneko2018macroscopic, kaneko2024constructing, kaneko2025universal}. 
However, explicitly addressing timescale separation requires a dynamical formulation. 
Deriving this principle as a dynamical process directly from the fundamental equations of population genetics remains an unresolved challenge.

In this paper, we formalize the general theory of the evolutionary Le Chatelier's principle. 
By utilizing the continuous-time Price equation \cite{price1970selection, price1972extension, frank1997price}, the foundational theorem of evolutionary dynamics, combined with the physical concept of timescale separation, we distinguish the fast relaxation of the expressed phenotype from the slow transgenerational evolution of a genotypic variable.
This approach allows us to construct a formulation of genetic assimilation that is independent of microscopic details.

\begin{figure*}[htbp]
\centering
\includegraphics[width=1.0\textwidth]{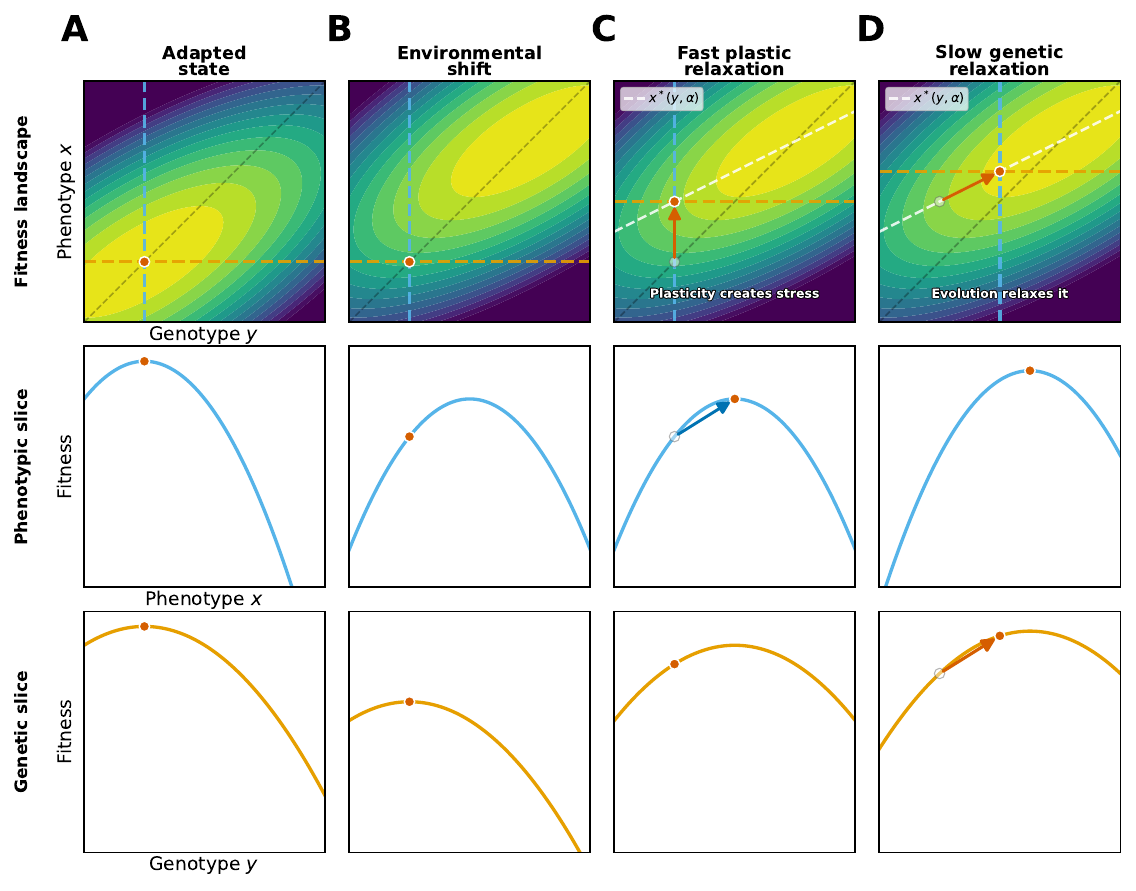}
\caption{
Dynamical relaxation picture of genetic assimilation governed by the evolutionary Le Chatelier's principle.
Columns show four successive stages: 
(A) the adapted state before the environmental change, 
(B) the environmental shift, 
(C) fast plastic relaxation of the phenotype, and 
(D) slow genetic relaxation.
The top row shows the fitness landscape \(F(x,y;\alpha)\) as a function of phenotype \(x\) and genotypic coordinate \(y\).
The middle and bottom rows show one-dimensional fitness slices along the phenotypic and genotypic coordinates, respectively.
Red circles indicate the current population state, and pale circles indicate the previous state before the movement shown by arrows.
Blue vertical and orange horizontal dashed lines in the top row indicate the fixed-\(y\) and fixed-\(x\) slices shown in the middle and bottom rows, respectively.
The white dashed curve in panels C and D denotes the constrained optimum \(x^*(y,\alpha)\).
In panel B, the environmental shift changes the fitness landscape, but at fixed phenotype and genotype it does not directly create a fitness-gradient component along the genotypic coordinate.
In panel C, fast plastic relaxation displaces the phenotype and creates internal stress through genotype--phenotype coupling.
In panel D, slow genetic evolution relaxes this stress, further moving the phenotype in the same direction as the initial plastic shift.
}
\label{fig:schematic}
\end{figure*}

Based on this framework, we address the dynamics of genetic assimilation (Fig.~\ref{fig:schematic}). 
When a sudden environmental change occurs, the environmental perturbation changes the fitness gradient along the fast phenotypic coordinate, but does not directly create a fitness-gradient component along the slow genotypic coordinate (Fig.~\ref{fig:schematic}B).
The phenotype therefore rapidly relaxes to a new constrained optimum, producing an immediate plastic displacement (Fig.~\ref{fig:schematic}C). 
Through genotype--phenotype coupling, this plastic displacement generates a selection gradient along the genotypic coordinate, or internal genetic stress, even though the environment does not directly generate that gradient.
Genetic assimilation then proceeds as a dynamical relaxation process in which the genotype evolves to resolve this self-generated stress (Fig.~\ref{fig:schematic}D). 
In short, plasticity creates stress, and evolution relaxes it.

To provide a physical picture of this general theory, we further apply our framework to an analytically tractable fitness landscape. 
In the explicit quadratic model below, genotype--phenotype coupling is identical to the plasticity cost, $F_{xy}=c$, which mathematically acts as a spring constant connecting the phenotype and genotype.

\section{Results}

\subsection{The Price equation with timescale separation}

We consider the continuous-time Price equation \cite{price1970selection, price1972extension, frank1997price}:
\begin{equation}
    \frac{d\overline{x}}{dt} = \mathrm{Cov}(x, F) + \overline{\left( \frac{dx}{dt} \right)}
    \label{eq:price_main}
\end{equation}
where $\overline{a}$ denotes the population mean of $a$, and $\mathrm{Cov}(a, b)$ represents the covariance between $a$ and $b$. 
$x$ is the phenotype of an individual, and $F$ is the fitness. 
The first term on the right-hand side indicates the change in the phenotype by selection, and the second term indicates the change driven by other factors (e.g., environmental changes).
Here, $t$ represents an evolutionary time scale across generations, while individual phenotypic changes (e.g., plastic responses) occur on a much shorter, intra-generational time scale.

We introduce the fitness function (or adaptive landscape \cite{wright1932roles}), $F(x, y; \alpha)$, given by the phenotype $x$, genotype $y$, and environment $\alpha$.
In addition, we apply three assumptions.
\\
\textbf{Assumption 1: No direct environmental selection on the genotypic variable}\\
We assume that $y$ is a slow inherited genotypic variable and that an environmental shift does not directly impose selection on this variable. In other words, at fixed phenotype and genotype, the environmental perturbation does not create a selection gradient along $y$.
\\
\textbf{Assumption 2: Timescale separation}\\
We assume a timescale separation between phenotype and genotype: the expressed phenotype relaxes much faster than the slow genotypic variable.
On the evolutionary timescale, this fast phenotypic relaxation is approximated as instantaneous, so that the phenotype is slaved to a constrained local optimum,
\begin{equation}
    \frac{\partial F(x^*, y; \alpha)}{\partial x} = 0
    \label{eq:qss_condition}
\end{equation}
where $x^* = x^*(y, \alpha)$. 
Thus, the plastic responses considered below correspond to local adaptive relaxation of the expressed phenotype under timescale separation.
\\
\textbf{Assumption 3: Local Quadratic Approximation}\\
We assume that the fitness function is locally approximated by a quadratic function. 
The second derivatives of the fitness function (Hessian components), such as $F_{xx} = \frac{\partial^2 F}{\partial x^2}$, $F_{yy} = \frac{\partial^2 F}{\partial y^2}$, $F_{xy} = \frac{\partial^2 F}{\partial x \partial y}$, and $F_{x\alpha} = \frac{\partial^2 F}{\partial x \partial \alpha}$, are constant and common across all individuals.
From Assumption 2, $x^*$ is on the peak of the fitness landscape, and thus $F_{xx} < 0$. 
Note that this assumption has been historically employed in quantitative genetics \cite{lande1979quantitative, lande1983measurement}.
In the local quadratic expansion, Assumption 1 is expressed as $F_{y\alpha}=0$, meaning that the environmental perturbation has no direct contribution to the fitness gradient along $y$.
By contrast, $F_{x\alpha}$ may be nonzero, allowing the environmental perturbation to induce a plastic response of the fast phenotypic variable.

By taking the total differential of Eq.~(\ref{eq:qss_condition}) (see SI Appendix for detailed derivations), we obtain the local response of the optimized phenotype as a linear function of $y$ and $\alpha$:
\begin{equation}
    x^*(y, \alpha) = -\frac{F_{xy}}{F_{xx}} y - \frac{F_{x\alpha}}{F_{xx}} \alpha + C
    \label{eq:pheno_x_star_exact}
\end{equation}
where $C$ is a constant of integration. 
Equation~(\ref{eq:pheno_x_star_exact}) has the formal structure of a local linear reaction norm: the coefficient $-F_{x\alpha}/F_{xx}$ represents the local plastic susceptibility, namely the environmental sensitivity of the expressed phenotype at fixed genotype.
This expression also identifies the phenotypic sensitivities to genetic and environmental perturbations. 
For fixed cross-couplings, a larger $\left| F_{xx} \right|$ makes the optimized phenotype less sensitive to changes in $y$ or $\alpha$, because the response coefficients scale with $1/F_{xx}$. 
This corresponds to a stiffer phenotypic response and connects to canalization-like robustness.

By substituting this solution into the Price equation (Eq.~\ref{eq:price_main}) and utilizing the linear properties of covariance, the total rate of phenotypic change is unified into the following form (see SI Appendix):
\begin{equation}
    \frac{d\overline{x}}{dt} = -\frac{F_{xy}}{F_{xx}} \frac{d\overline{y}}{dt} - \frac{F_{x\alpha}}{F_{xx}} \frac{d\alpha}{dt}
    \label{eq:price_unified}
\end{equation}
This equation demonstrates that the phenotypic change can be explicitly decoupled into two distinct biological contributions: the slow dynamics driven by genetic evolution ($\frac{d\overline{x}_\mathrm{evo}}{dt}$) and the fast dynamics due to phenotypic plasticity ($\frac{d\overline{x}_\mathrm{plas}}{dt}$).
We define them respectively as:
\begin{equation}
    \frac{d\overline{x}_\mathrm{evo}}{dt} = -\frac{F_{xy}}{F_{xx}} \frac{d\overline{y}}{dt}, \quad \frac{d\overline{x}_\mathrm{plas}}{dt} = -\frac{F_{x\alpha}}{F_{xx}} \frac{d\alpha}{dt}
    \label{eq:pheno_decomp}
\end{equation}
Thus, the Price equation is rewritten as the sum of these two components:
\begin{equation}
    \frac{d\overline{x}}{dt} = \frac{d\overline{x}_\mathrm{evo}}{dt} + \frac{d\overline{x}_\mathrm{plas}}{dt}
    \label{eq:price_unified_decomp}
\end{equation}

To explicitly determine the evolutionary dynamics of the slow variable $y$ and further expand the evolutionary component, we introduce a standard assumption from quantitative genetics \cite{falconer1996introduction, crow1970introduction, lynch1998genetics, burger2000mathematical}. 
\\
\textbf{Assumption 4: Symmetric distribution of genotype}\\
We assume that $y$ has a symmetric distribution, and thus odd-ordered moments of $y$ are zero.

Since $y$ is a slow inherited genotypic variable, its evolutionary dynamics follow the Price equation with the selection term, $\frac{d\overline{y}}{dt} = \mathrm{Cov}(y, F)$.
By evaluating this covariance term through a quadratic expansion of the fitness function under Assumptions 3 and 4 (see SI Appendix), the evolutionary rate defined in Eq.~(\ref{eq:pheno_decomp}) is derived as:
\begin{equation}
    \frac{d\overline{x}_\mathrm{evo}}{dt} = -G_{yy} \frac{F_{xy}}{F_{xx}} \frac{\partial F}{\partial \overline{y}}
    \label{eq:price_evo_Lande}
\end{equation}
where $G_{yy}$ is the variance of $y$.
This formulation is essentially equivalent to Lande's equation \cite{lande1979quantitative}, which generalizes Fisher's fundamental theorem of natural selection \cite{fisher1930genetical} to arbitrary phenotypic traits.

\subsection{Correspondence between the Price equation and Le Chatelier's principle}

We consider a situation where the environment shifts by $\Delta \alpha$ at $t = 0$. 
The time evolution of $\alpha$ is represented as $\frac{d \alpha}{dt} = \Delta \alpha \delta(t)$. 
We assume $y$ is initially optimized before the environmental shift. 
From the stability condition, $\left. \frac{\partial F}{\partial \overline{y}} \right|_{t=0^-} = 0$, $F_{yy} < 0$, and $\det H = F_{xx} F_{yy} - F_{xy}^2 > 0$ are satisfied, where $H$ is the Hessian matrix of the fitness landscape.

Because $\overline{y}$ is a slow variable driven by genetic evolution, its change over an infinitesimal time is zero. 
The initial phenotypic response, $\Delta \overline{x}_\mathrm{plas}$, is entirely determined by the plastic component (see SI Appendix):
\begin{equation}
    \Delta \overline{x}_\mathrm{plas} = -\frac{F_{x\alpha}}{F_{xx}} \Delta \alpha
    \label{eq:pheno_short_term_resp}
\end{equation}

Although the environmental shift does not directly generate a genetic gradient, the plastic displacement does. 
Expanding the genetic gradient immediately after the shift gives $\left. \frac{\partial F}{\partial \overline{y}} \right|_{t=0^+} = F_{xy}\Delta \overline{x}_{\rm plas}$ (SI Appendix), leading to the initial evolutionary velocity
\begin{equation}
    \left. \frac{d\overline{x}_\mathrm{evo}}{dt} \right|_{t=0^+} = -G_{yy} \frac{F_{xy}^2}{F_{xx}} \Delta \overline{x}_\mathrm{plas}
    \label{eq:pheno_initial_velocity}
\end{equation}
Since $G_{yy} > 0$ and $F_{xx} < 0$, the coefficient $-G_{yy} \frac{F_{xy}^2}{F_{xx}}$ is positive. 
Thus, the initial genetic component of the phenotypic response is in the same direction as the initial plastic shift. 
In this way, genotype--phenotype coupling converts the plastic displacement into the internal stress \(F_{xy}\Delta \overline{x}_{\rm plas}\) that drives assimilation.
For a fixed plastic displacement, a larger $|F_{xx}|$ slows the initial evolutionary response.

At the final steady state ($t \rightarrow \infty$), genetic evolution ceases, satisfying $\frac{\partial F}{\partial \overline{y}} (\infty) = 0$. Solving this condition (see SI Appendix) yields the total change in genotype:
\begin{equation}
    \Delta \overline{y}(\infty) = - \frac{F_{xx} F_{xy}}{\det H} \Delta \overline{x}_\mathrm{plas}
    \label{eq:geno_steady_state}
\end{equation}
Eq.~(\ref{eq:geno_steady_state}) provides a dynamical interpretation of the evolutionary Le Chatelier's principle. 
Rewriting Eq.~(\ref{eq:geno_steady_state}) as $- \frac{\det H}{F_{xx}} \Delta \overline{y}(\infty) = F_{xy} \Delta \overline{x}_\mathrm{plas}$ shows the force balance. The right-hand side represents the internal stress generated by the plastic deformation, while the left-hand side represents the restoring force generated by the displacement of the slow variable $y$. 
Thus, genetic assimilation functions as a relaxation process to cancel out this internal stress.

The total long-time change in the phenotype, $\Delta \overline{x}_\mathrm{total} = \Delta \overline{x}_\mathrm{evo}(\infty) + \Delta \overline{x}_\mathrm{plas}$, is derived as (see SI Appendix):
\begin{equation}
    \Delta \overline{x}_\mathrm{total} = \left(1 + \frac{F_{xy}^2}{\det H} \right) \Delta \overline{x}_\mathrm{plas}
    \label{eq:le_chatelier_eq}
\end{equation}
Since $\det H > 0$, the following inequality is satisfied:
\begin{equation}
    \left| \Delta \overline{x}_\mathrm{total} \right| \geq \left| \Delta \overline{x}_\mathrm{plas} \right| 
    \label{eq:le_chatelier_ineq}
\end{equation}
This inequality corresponds to the phenomenological manifestation of Le Chatelier's principle. 
The total phenotypic change caused by both phenotypic plasticity and evolution is always in the same direction and equal to or greater than that caused by phenotypic plasticity alone. 
Therefore, genetic assimilation enhances the phenotypic change initially triggered by phenotypic plasticity.

\subsection{Exponential decay and the timescale of genetic assimilation}

We examine how genetic assimilation changes over time. 
The evolutionary process of the genotype is driven by the Price equation $\frac{d\overline{y}}{dt} = G_{yy} \frac{\partial F}{\partial \overline{y}}$. 
By evaluating the slope $\frac{\partial F}{\partial \overline{y}}$ at an arbitrary time (see SI Appendix), the evolutionary dynamics can be rewritten as a linear differential equation:
\begin{equation}
    \frac{d}{dt} \Delta \overline{y} (t) = - \frac{1}{\tau} \left( \Delta \overline{y} (t) - \Delta \overline{y}(\infty) \right)
    \label{eq:geno_dynamics}
\end{equation}
where $\tau$ is the characteristic relaxation time for genetic assimilation, defined as:
\begin{equation}
    \tau = - \frac{F_{xx}}{G_{yy} \det H}
    \label{eq:tau_def}
\end{equation}
Eq.~(\ref{eq:geno_dynamics}) indicates that the genotype $\overline{y}$ relaxes toward its evolutionary destination $\Delta \overline{y}(\infty)$ to resolve the internal stress induced by the initial plastic response. 
Since the initial genetic displacement is zero ($\Delta \overline{y} (0^+) = 0$), the analytical solution is an exponential decay function:
\begin{equation}
    \Delta \overline{y} (t) = \Delta \overline{y}(\infty) \left( 1 - e^{-\frac{t}{\tau}} \right)
    \label{eq:geno_solution}
\end{equation}
This implies that the internal stress driving genetic assimilation decays exponentially over time.
Eq.~(\ref{eq:tau_def}) shows that the relaxation time is controlled by the local phenotypic stiffness $|F_{xx}|$, together with the remaining Hessian structure, $\det H$, and the genetic variance $G_{yy}$.
Thus, the same stiffness also enters the timescale over which the internally generated stress is relaxed.

To determine the variance of $y$, $G_{yy}$, we strengthen Assumption 4 and introduce a new assumption.
\\
\textbf{Assumption 4': Gaussian distribution of genotype}\\
We assume that the genotypic distribution $P(y)$ in the population follows a Gaussian distribution with variance $G_{yy}$ \cite{falconer1996introduction, crow1970introduction, lynch1998genetics, burger2000mathematical}.
\\
\textbf{Assumption 5: Mutation-selection balance}\\
We assume that the genetic variance is maintained at a steady state by a dynamic equilibrium between the reduction of variance due to stabilizing selection and the continuous input of novel genetic variation via mutation ($V_m \ll 1$). 
The steady-state balance is $G_{yy} - G_{yy}' = V_m$, where $G_{yy}'$ is the genetic variance after selection in a single generation.

We evaluate the effective fitness landscape experienced by the slow genotypic variable $y$ when the fast phenotype $x$ is constantly optimized.
The effective curvature, representing the strength of stabilizing selection on $y$, is derived as $\kappa = -\frac{\det H}{F_{xx}}$ (see SI Appendix). 
Under Assumption 4', applying this selection pressure gives the updated variance $G'_{yy} = \frac{G_{yy}}{1 + \kappa G_{yy}}$ (SI Appendix). 

Applying Assumption 5 (Mutation-selection balance) \cite{kimura1965stochastic, burger2000mathematical, lande1975maintenance, lynch1998genetics}, we solve for the steady-state genetic variance. 
For $V_m \ll 1$, we obtain an approximated variance (SI Appendix):
\begin{equation}
    G_{yy} \simeq \sqrt{\frac{V_m}{\kappa}} = \sqrt{-\frac{V_{m} F_{xx}}{\det H}}
    \label{eq:var_approx_solution}
\end{equation}
Substituting Eq.~(\ref{eq:var_approx_solution}) into Eq.~(\ref{eq:tau_def}) gives the characteristic timescale solely characterized by the macroscopic parameters of the landscape and mutation:
\begin{equation}
    \tau = \sqrt{- \frac{F_{xx}}{V_{m} \det H}}
    \label{eq:tau_balance}
\end{equation}
This indicates that the speed of genetic assimilation is determined by the curvature of the fitness landscape ($F_{xx}$ and $\det H$) and the mutational rate ($V_m$). 
As demonstrated in the specific example below, this timescale provides biological insights into how developmental constraints and selection pressures dictate the speed of evolution.

\subsection{Example: A quadratic fitness landscape with a selection pressure and a cost of phenotypic plasticity}

We apply our theoretical framework to a specific quantitative genetic model: a quadratic fitness landscape with a selection pressure and a cost of phenotypic plasticity. 
The fitness function $F(x, y; \alpha)$ is defined as:
\begin{equation}
    F(x, y; \alpha) = -\frac{s}{2} (x - \alpha)^2 - \frac{c}{2}(x-y)^2
    \label{eq:ex_fitness}
\end{equation}
The first term represents stabilizing selection toward the environmental optimum $\alpha$, where $s > 0$ is the strength of the selection pressure. 
The second term represents the cost of phenotypic plasticity, penalizing the deviation of the phenotype $x$ from the baseline genotype $y$, where $c > 0$ is the magnitude of this cost.

The second derivatives (Hessian components) of this fitness landscape are:
\begin{equation}
    F_{xx} = -(s+c), \quad F_{yy} = -c, \quad F_{xy} = c, \quad F_{x\alpha} = s
    \label{eq:ex_derivatives}
\end{equation}
The determinant of the Hessian matrix, $\det H$, is calculated as:
\begin{equation}
    \det H = F_{xx} F_{yy} - F_{xy}^2 = -(s+c)(-c) - c^2 = sc
    \label{eq:ex_detH}
\end{equation}
The stability of the fitness landscape ($\det H > 0$) requires both environmental selection pressure ($s > 0$) and a cost of plasticity ($c > 0$). 
If either is zero, $\det H = 0$, meaning the fitness landscape forms a flat ridge, and the uniqueness of the optimal state is lost.

We calculate the initial fast response of the phenotype ($\Delta \overline{x}_\mathrm{plas}$) to an environmental shift $\Delta \alpha$. 
Substituting Eq.~(\ref{eq:ex_derivatives}) into Eq.~(\ref{eq:pheno_short_term_resp}) gives:
\begin{equation}
    \Delta \overline{x}_\mathrm{plas} = -\frac{s}{-(s+c)} \Delta \alpha = \frac{s}{s+c} \Delta \alpha
    \label{eq:ex_x_plas}
\end{equation}
This indicates an incomplete plastic adaptation; the initial phenotypic shift falls short of the environmental change $\Delta \alpha$ due to the plasticity cost $c$. 

As discussed in the previous section regarding the force balance (see the text following Eq.~\ref{eq:geno_steady_state}), the internal stress initially driving genetic assimilation is represented by $F_{xy} \Delta \overline{x}_\mathrm{plas}$. 
In this model, since $F_{xy}=c$, the internal stress is explicitly expressed as $c \Delta \overline{x}_\mathrm{plas}$. 
The plasticity cost $c$ mathematically acts as a spring constant connecting the genotype and phenotype. 
Genetic assimilation is the dynamic process of relaxing the elastic energy stored in this spring.

Substituting Eqs.~(\ref{eq:ex_derivatives}) and (\ref{eq:ex_detH}) into Eq.~(\ref{eq:geno_steady_state}) yields the total evolutionary change of the genotype ($\Delta \overline{y}(\infty)$) after sufficient time has passed:
\begin{equation}
    \Delta \overline{y}(\infty) = - \frac{-(s+c)c}{sc} \left( \frac{s}{s+c} \Delta \alpha \right) = \Delta \alpha
    \label{eq:ex_y_evo}
\end{equation}
By substituting the parameters into Eq.~(\ref{eq:le_chatelier_eq}), the total phenotypic change ($\Delta \overline{x}_\mathrm{total}$) is:
\begin{equation}
    \Delta \overline{x}_\mathrm{total} = \left( 1 + \frac{c^2}{sc} \right) \left( \frac{s}{s+c} \Delta \alpha \right) = \Delta \alpha
    \label{eq:ex_x_total}
\end{equation}

Since $s > 0$ and $c > 0$, the inequality $|\Delta \overline{x}_\mathrm{total}| = |\Delta \alpha| > |\Delta \overline{x}_\mathrm{plas}| = \frac{s}{s+c}|\Delta \alpha|$ holds. 
This confirms the phenomenological manifestation of the evolutionary Le Chatelier's principle. Comparing Eq.~(\ref{eq:ex_y_evo}) and Eq.~(\ref{eq:ex_x_total}) reveals that the final states of the phenotype and genotype are aligned ($\Delta \overline{x}_\mathrm{total} = \Delta \overline{y}(\infty) = \Delta \alpha$). 
This indicates that genetic assimilation proceeds to eliminate the plasticity cost.

We evaluate the relaxation time $\tau$, which governs the speed of this genetic assimilation. Substituting Eq.~(\ref{eq:ex_derivatives}) and Eq.~(\ref{eq:ex_detH}) into Eq.~(\ref{eq:tau_def}), we obtain:
\begin{equation}
    \tau = - \frac{F_{xx}}{G_{yy} \det H} = \frac{s+c}{G_{yy} s c} = \frac{1}{G_{yy}} \left( \frac{1}{s} + \frac{1}{c} \right)
    \label{eq:ex_tau}
\end{equation}
Under the condition where a balance between selection and mutation is satisfied (Eq.~\ref{eq:tau_balance}), this becomes:
\begin{equation}
    \tau = \sqrt{ \frac{1}{V_m} \left( \frac{1}{s} + \frac{1}{c} \right)}
    \label{eq:ex_tau_balance}
\end{equation}
For fixed genetic variance $G_{yy}$, Eq.~(\ref{eq:ex_tau}) shows that the time required for genetic assimilation is proportional to the sum of the inverses of the selection pressure ($s$) and the plasticity cost ($c$). 
Under mutation--selection balance, Eq.~(\ref{eq:ex_tau_balance}) instead shows that the relaxation time scales with the square root of this sum. 
In both cases, if either the environmental selection pressure is absent ($s \to 0$) or the plasticity is cost-free ($c \to 0$), the relaxation time diverges ($\tau \to \infty$). 
While phenotypic plasticity allows organisms to survive sudden environmental changes, the cost of maintaining this plasticity drives the subsequent genetic assimilation.

\begin{figure*}[htbp]
\centering
\includegraphics[width=0.78\textwidth]{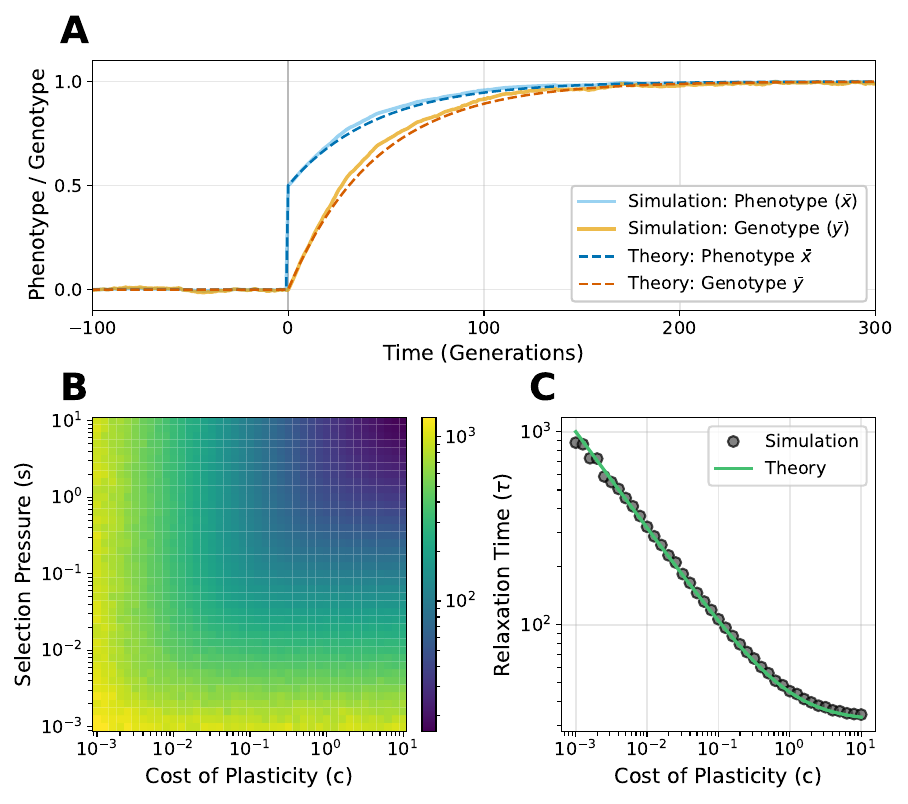}
\caption{
Individual-based simulation of genetic assimilation on the quadratic fitness landscape.
(A) Time series of the mean phenotype \(\bar{x}\) and mean genotype \(\bar{y}\) before and after an environmental shift \(\Delta\alpha=1\) at \(t=0\).
Solid lines show a representative single replicate for a population of \(N=10{,}000\) with \(s=1.0\) and \(c=1.0\), and dashed lines show the analytical predictions with no fitted parameters.
(B) Relaxation time \(\tau\) obtained from parameter sweeps over the selection pressure \(s\) and the cost of plasticity \(c\), averaged over 100 independent replicates for each parameter combination.
(C) Cross-section of the relaxation time at fixed selection pressure \(s=1.0\).
Points show mean simulation results averaged over 100 independent replicates, and the solid line shows the theoretical prediction from Eq.~(25).
}
\label{fig:simulation}
\end{figure*}

To validate the theoretical framework, we performed an individual-based simulation of a finite population evolving on this quadratic fitness landscape (Fig.~\ref{fig:simulation}). 
As shown in Fig.~\ref{fig:simulation}A, when the environment instantaneously shifts at $t=0$, the mean phenotype exhibits an immediate plastic jump, partially adapting to the new environment. 
Subsequently, the mean genotype slowly evolves to close the phenotype--genotype gap. The theoretical predictions (dashed lines) capture this microscopic population dynamics. 

We evaluated the characteristic relaxation time $\tau$ across a range of selection pressures ($s$) and plasticity costs ($c$). 
Fig.~\ref{fig:simulation}B demonstrates that the relaxation time becomes long when the cost of plasticity is small ($c \to 0$) or the selection pressure is weak ($s \to 0$). 
The theoretically predicted $\tau$ shows quantitative agreement with the empirical relaxation times observed in the simulation (Fig.~\ref{fig:simulation}C). 
These numerical results are consistent with the theoretical prediction that, in the quadratic model, the plasticity cost physically acts as the internal spring constant driving genetic assimilation.

\section{Discussion}
In this paper, we formulated a macroscopic framework for genetic assimilation by bridging the Price equation with the physical concept of timescale separation. 
Our formulation reveals that the evolutionary trajectory following an environmental shift is governed by the evolutionary Le Chatelier's principle. 
Genetic assimilation is fundamentally a dynamical relaxation process: the initial plastic displacement generates an internal stress through genotype--phenotype coupling, and subsequent genetic evolution proceeds to resolve this stress.
This dynamical picture demonstrates that genetic assimilation is a natural physical consequence of evolving systems undergoing timescale separation, independent of microscopic genetic details.

While previous models describe genetic assimilation through specific genetic architectures \cite{lande1979quantitative, lande2009adaptation, chevin2010adaptation} or explicit biophysical and developmental constraints \cite{ancel2000plasticity, espinosa2011phenotypic, raju2023theoretical}, our formulation provides a phenomenological counterpart. 
By coarse-graining these microscopic details into a generalized potential landscape under the assumption of timescale separation, our framework complements these mechanistic approaches. 
Rather than detailing how specific traits are mechanistically encoded, this phenomenological perspective explains why genetic assimilation follows a universal macroscopic relaxation law.

Equation~(\ref{eq:pheno_x_star_exact}) has the formal structure of a local linear reaction norm and is therefore closely related to quantitative-genetic models of plasticity, particularly Lande's model of adaptation to an extraordinary environment \cite{lande2009adaptation}. 
The distinction is that the present theory does not postulate the reaction norm as a microscopic genetic architecture. 
Instead, the reaction-norm-like slope emerges as a local susceptibility after eliminating the fast phenotypic variable through $F_x=0$.
Thus, Lande's model can be viewed as a microscopic realization in which this susceptibility itself can evolve, whereas the present theory treats the susceptibility as locally fixed and focuses on the macroscopic relaxation law generated by timescale separation.

A biological implication of our framework lies in the role of genotype--phenotype coupling. 
In the general theory, the internal stress driving assimilation is $F_{xy}\Delta \overline{x}_\mathrm{plas}$. 
In the quadratic model, this coupling is identical to the plasticity cost, $F_{xy}=c$, so the cost acts as a spring constant connecting phenotype and genotype. 
This explains why a cost that is conventionally viewed as an evolutionary constraint can also serve as the restoring force that drives genetic assimilation.

The limit $c\to0^+$ should be interpreted as a singular limit of the stable quadratic model. 
The restoring force between genotype and phenotype vanishes, $\det H=sc$ approaches zero, and the relaxation time diverges. 
Thus, in the ideal limit of cost-free, perfectly adaptive plasticity, genetic assimilation effectively stalls.

This relationship provides a testable prediction for experimental evolution. 
In the quadratic model, the plasticity cost \(c\) sets the genotype--phenotype restoring force. 
If developmental noise is comparable across genotypes and environments, a weaker restoring force should be reflected in larger phenotypic fluctuations among genetically identical individuals. 
Thus, clonal phenotypic fluctuations provide a practical way to infer the strength of the local restoring force.
Our theory predicts that a population exhibiting larger clonal phenotypic fluctuations will require a longer evolutionary timescale for genetic assimilation. 
This counter-intuitive prediction provides an avenue to test the evolutionary Le Chatelier's principle in laboratory experiments, demonstrating that greater clonal phenotypic variation slows genetic evolution because of a weaker restoring force.

While our explicit example assumed a one-dimensional trait, extending this theoretical framework to multidimensional phenotypic and genotypic spaces remains an important direction for future research. Real biological systems typically involve pleiotropy and multidimensional fitness landscapes. 
By replacing the scalar variance with a genetic covariance matrix and the second derivative of fitness with a Hessian matrix, this formulation could be generalized to describe a multidimensional evolutionary Le Chatelier's principle. 

Beyond simple quadratic landscapes, exploring evolution on complex, rugged fitness landscapes represents another direction for future research. 
In highly rugged landscapes, the quantitative relationships derived here might be complicated by local optima and rare evolutionary events, such as valley crossing \cite{hatakeyama2024evolutionary}, required for global adaptation. 
However, the core qualitative picture would remain valid: the initial plastic displacement creates an internal stress through genotype--phenotype coupling, driving subsequent evolution.

A further extension concerns the Baldwin effect. 
The Baldwin effect is historically associated with learning and other forms of ontogenetic accommodation \cite{baldwin1896new, simpson1953baldwin}. 
It is related to genetic assimilation because both involve plastic responses that influence subsequent evolution \cite{crispo2007baldwin}, but the two concepts should not be conflated \cite{loison2020disentangling}. 
In the present framework, the natural extension would be to introduce an additional fast ontogenetic process, with learning providing the clearest example. 
Such a process would not transmit acquired behavior to the genome; rather, it would change how genotypes are evaluated by selection. 
Classic models show that learning can allow partially suitable genotypes to survive and reproduce, making otherwise hidden genetic variation visible to natural selection \cite{hinton1987how}. 
In physical terms, this additional fast process reshapes the effective fitness landscape experienced by the slower phenotypic and genotypic variables.
A future extension could coarse-grain over this fast accommodative process before deriving the responses of the slower phenotypic and genotypic variables.
Such a multiple-timescale formulation could clarify when fast ontogenetic accommodation accelerates genetic adaptation and when its costs or limits favor reduced reliance on the fast response.

Our results establish a formal link between evolutionary dynamics and macroscopic physics. 
Just as Fisher's fundamental theorem mirrors the second law of thermodynamics by dictating the monotonic increase of mean fitness \cite{fisher1930genetical}, the evolutionary Le Chatelier's principle mathematically dictates how an evolving system macroscopically responds to external perturbations. 
By unifying phenotypic plasticity and genetic evolution under this universal framework, we anticipate that this macroscopic perspective will stimulate further theoretical and empirical explorations into the physical nature of biological evolution.

\section*{Materials and Methods}
\subsection*{Individual-based simulation}
We simulated a Wright-Fisher population of $N=10,000$ individuals with discrete, non-overlapping generations. 
Each individual $i$ in the population is characterized by a continuous genotype $y_i$ and a phenotype $x_i$.
The absolute fitness of an individual is given by $W(x_i,y_i)=\exp[F(x_i,y_i;\alpha)]$, where $F$ is the quadratic fitness function defined in Eq.~(\ref{eq:ex_fitness}).

Following the assumption of timescale separation, the actual phenotype $x_i$ expressed by an individual is assumed to be constantly kept at its optimal plastic response: $x_i=x^*(y_i,\alpha)=\frac{s\alpha+cy_i}{s+c}$.
In each generation, $N$ parents are selected to produce the next generation with probabilities proportional to their absolute fitness $W(x_i,y_i)$.
The offspring inherit their parent's genotype with a mutational perturbation: $y_i'=y_i+\xi_m$, where the mutational effect $\xi_m$ is drawn from a normal distribution with mean zero and variance $V_m=0.001$.

The population was initialized with $y_i=0$ for all individuals and allowed to evolve for a burn-in period of 2,000 generations under a constant environment ($\alpha=0$) to reach a mutation-selection balance.
At $t=0$, the environment was instantaneously shifted to $\alpha=1$.
For the time-series analysis (Fig.~\ref{fig:simulation}A), a single replicate run with $s=1.0$ and $c=1.0$ was performed to demonstrate the representative evolutionary trajectory. 
For the parameter sweeps (Fig.~\ref{fig:simulation}B and C), the parameters $s$ and $c$ were independently varied from $10^{-3}$ to $10^{1}$, uniformly divided into 41 points on a logarithmic scale.
The empirical relaxation time $\tau$ was numerically defined as the number of generations required for the mean genotype $\bar{y}$ to reach $1-e^{-1}\simeq63.2\%$ of its theoretically predicted total displacement $\Delta\bar{y}(\infty)$. 
The presented sweep data were averaged over 100 independent replicate runs for each parameter combination.

\begin{acknowledgments}
I thank Chikara Furusawa for helpful discussions, and Kunihiko Kaneko and Archishman Raju for helpful discussions and critical reading of the manuscript. 
This work was supported by JSPS KAKENHI Grant Numbers 26K00057 and 26K00061.
\end{acknowledgments}

\bibliography{refs}

\clearpage
\onecolumngrid
\setcounter{equation}{0}
\renewcommand{\theequation}{S\arabic{equation}}
\renewcommand{\theHequation}{S\arabic{equation}}

\section*{Supplemental Material}

\section*{S1. Derivation of the Price equation under timescale separation}

This section provides the detailed derivations of the Price-equation formulation and the local phenotypic response presented in the main text.

\subsection*{Derivation of the local response of the optimized phenotype}
From Assumption 2, the phenotype is kept at the optimal state $x^*$, satisfying $\partial F / \partial x = 0$. By taking the total differential of this condition, we obtain:
\begin{align}
    d\left( \frac{\partial F}{\partial x} \right) &= \frac{\partial^2 F}{\partial x^2} dx^* + \frac{\partial^2 F}{\partial x \partial y} dy + \frac{\partial^2 F}{\partial x \partial \alpha} d\alpha \nonumber \\
    &= F_{xx} dx^* + F_{xy} dy + F_{x\alpha} d\alpha \nonumber \\
    &= 0
    \label{eq:S_fitness_diff}
\end{align}
By rearranging this, we obtain the local response of the optimized phenotype:
\begin{equation}
    dx^* = -\frac{F_{xy}}{F_{xx}} dy - \frac{F_{x\alpha}}{F_{xx}} d\alpha
    \label{eq:S_pheno_dx_star}
\end{equation}
Because the Hessian components are constant under Assumption 3, this differential equation can be directly integrated by treating $y$ and $\alpha$ as independent coordinates:
\begin{equation}
    x^*(y, \alpha) = -\frac{F_{xy}}{F_{xx}} y - \frac{F_{x\alpha}}{F_{xx}} \alpha + C
    \label{eq:S_pheno_x_star_exact}
\end{equation}
where $C$ is a constant of integration. The partial derivatives are determined as constants:
\begin{equation}
    \frac{\partial x^*}{\partial y} = -\frac{F_{xy}}{F_{xx}}, \quad \frac{\partial x^*}{\partial \alpha} = -\frac{F_{x\alpha}}{F_{xx}}
    \label{eq:S_pheno_partial}
\end{equation}

\subsection*{Decomposition of the Price equation}
We evaluate the second term on the right-hand side of the continuous-time Price equation; 
\begin{equation}
    \frac{d\overline{x}}{dt} = \mathrm{Cov}(x, F) + \overline{\left( \frac{dx}{dt} \right)}
    \label{eq:S_price_main}
\end{equation}
Since the environment $\alpha$ changes over time, the rate of phenotypic change due to the environment is:
\begin{align}
    \frac{dx^*}{dt} &= \frac{\partial x^*}{\partial \alpha} \frac{d\alpha}{dt} \nonumber \\
    &= -\frac{F_{x\alpha}}{F_{xx}} \frac{d\alpha}{dt}
    \label{eq:S_pheno_dx_dt}
\end{align}
Because $F_{xx}$, $F_{x\alpha}$, and $d\alpha/dt$ are common to every individual, its population average is simply:
\begin{equation}
    \overline{\left( \frac{dx}{dt} \right)} = -\frac{F_{x\alpha}}{F_{xx}} \frac{d\alpha}{dt}
    \label{eq:S_pheno_mean_dx_dt}
\end{equation}

Next, we examine the first term of the Price equation, $\mathrm{Cov}(x, F)$. When calculating the covariance at a specific time $t$, the macroscopic environmental parameter $\alpha$ is a global state common to all individuals ($\mathrm{Cov}(\alpha, F) = 0$). By applying the linear property of covariance to the solution of $x^*$, we obtain:
\begin{align}
    \mathrm{Cov}(x, F) &= \mathrm{Cov}(x^*, F) \nonumber \\
    &= \mathrm{Cov}\left( -\frac{F_{xy}}{F_{xx}} y - \frac{F_{x\alpha}}{F_{xx}} \alpha + C, F \right) \nonumber \\
    &= -\frac{F_{xy}}{F_{xx}} \mathrm{Cov}(y, F)
    \label{eq:S_cov_x_expand}
\end{align}
Because $y$ is a slow inherited genotypic variable, its evolutionary dynamics follow the Price equation with the selection term:
\begin{equation}
    \frac{d\overline{y}}{dt} = \mathrm{Cov}(y, F)
    \label{eq:S_price_y}
\end{equation}
Substituting this into Eq.~(\ref{eq:S_cov_x_expand}) gives:
\begin{equation}
    \mathrm{Cov}(x, F) = -\frac{F_{xy}}{F_{xx}} \frac{d\overline{y}}{dt}
    \label{eq:S_cov_sub}
\end{equation}
By substituting Eqs.~(\ref{eq:S_pheno_mean_dx_dt}) and (\ref{eq:S_cov_sub}) into the Price equation, it is unified into the following form:
\begin{equation}
    \frac{d\overline{x}}{dt} = -\frac{F_{xy}}{F_{xx}} \frac{d\overline{y}}{dt} - \frac{F_{x\alpha}}{F_{xx}} \frac{d\alpha}{dt}
    \label{eq:S_price_unified}
\end{equation}
We decompose the total rate of phenotypic change into two biological contributions: the slow dynamics driven by genetic evolution ($\frac{d\overline{x}_\mathrm{evo}}{dt}$) and the fast dynamics due to phenotypic plasticity ($\frac{d\overline{x}_\mathrm{plas}}{dt}$). We define them respectively as:
\begin{equation}
    \frac{d\overline{x}_\mathrm{evo}}{dt} = -\frac{F_{xy}}{F_{xx}} \frac{d\overline{y}}{dt}, \quad \frac{d\overline{x}_\mathrm{plas}}{dt} = -\frac{F_{x\alpha}}{F_{xx}} \frac{d\alpha}{dt}
    \label{eq:S_pheno_decomp}
\end{equation}
Thus, the Price equation is decomposed into the sum of these two components:
\begin{equation}
    \frac{d\overline{x}}{dt} = \frac{d\overline{x}_\mathrm{evo}}{dt} + \frac{d\overline{x}_\mathrm{plas}}{dt}
    \label{eq:S_price_unified_decomp}
\end{equation}

\subsection*{Derivation of the evolutionary rate}
To explicitly evaluate $\mathrm{Cov}(y, F)$, we expand the fitness function up to the quadratic term around the population mean:
\begin{align}
    F(x, y; \alpha) &= F(\overline{x}, \overline{y}; \alpha) + \frac{\partial F}{\partial \overline{x}} \left( x - \overline{x} \right) + \frac{\partial F}{\partial \overline{y}} \left( y - \overline{y} \right) \nonumber \\ 
    &+ \frac{1}{2} F_{xx} \left( x - \overline{x} \right)^2 + F_{xy} \left( x - \overline{x} \right) \left( y - \overline{y} \right) + \frac{1}{2} F_{yy} \left( y - \overline{y} \right)^2
    \label{eq:S_fitness_expand}
\end{align}
In the covariance calculation, all terms involving only $\alpha$ or cross-terms with $\alpha$ are zero because the variance of $\alpha$ within the population is zero. From the optimal condition, the term proportional to $\frac{\partial F}{\partial x}$ is zero. Using the linear relation $x-\overline{x} = - (F_{xy}/F_{xx}) ( y - \overline{y} )$, we calculate $\mathrm{Cov}(y, F)$:
\begin{align}
    \mathrm{Cov}(y, F) &= \overline{\left[ \left( y - \overline{y} \right) F(x,y;\alpha) \right]} \nonumber \\
    & = F(\overline{x}, \overline{y}; \alpha) \overline{ \left( y - \overline{y} \right)} + \frac{\partial F}{\partial \overline{y}} \overline{\left( y - \overline{y} \right)^2} + \frac{1}{2} \left( F_{yy} - \frac{F_{xy}^2}{F_{xx}} \right) \overline{\left( y - \overline{y} \right)^3} \nonumber \\
    &= \frac{\partial F}{\partial \overline{y}} \overline{\left( y - \overline{y} \right)^2} \nonumber \\
    &= G_{yy} \frac{\partial F}{\partial \overline{y}} = \frac{d \overline{y}}{dt}
    \label{eq:S_cov_y_expand}
\end{align}
where we used the definition $\overline{(y-\overline{y})} = 0$, and under Assumption 4 (symmetric distribution), the third central moment $\overline{(y-\overline{y})^3} = 0$. Substituting this result into the evolutionary term defined in Eq.~(\ref{eq:S_pheno_decomp}) yields the evolutionary rate:
\begin{equation}
    \frac{d\overline{x}_\mathrm{evo}}{dt} = -G_{yy} \frac{F_{xy}}{F_{xx}} \frac{\partial F}{\partial \overline{y}}
    \label{eq:S_price_evo_Lande}
\end{equation}

\section*{S2. Derivation of the dynamic response and Le Chatelier's principle}

This section provides the detailed derivations of the dynamic response and the evolutionary Le Chatelier relation presented in the main text.

We consider an environmental shift $\Delta \alpha$ at $t = 0$, defined as $\frac{d \alpha}{dt} = \Delta \alpha \delta(t)$. Prior to the shift, $y$ is optimized, satisfying $\left. \frac{\partial F}{\partial \overline{y}} \right|_{t=0^-} = 0$, $F_{yy} < 0$, and $\det H = F_{xx} F_{yy} - F_{xy}^2 > 0$.

\subsection*{Short-time response}
We calculate the fast response of $\overline{x}$ in the short-time limit ($\epsilon \rightarrow 0$). Because $\overline{y}$ is a slow variable, its change over an infinitesimal time is zero ($\int_{-\epsilon}^{+\epsilon} \frac{d\overline{x}_\mathrm{evo}}{dt} dt = 0$). Integrating the plastic component yields the initial phenotypic response $\Delta \overline{x}_\mathrm{plas}$:
\begin{align}
    \Delta \overline{x}_\mathrm{plas} &= \int^{+\epsilon}_{-\epsilon} \frac{d\overline{x}_\mathrm{plas}}{dt} dt \nonumber \\
    &= - \int^{+\epsilon}_{-\epsilon}\frac{F_{x\alpha}}{F_{xx}} \Delta \alpha \delta(t) dt \nonumber \\
    &= -\frac{F_{x\alpha}}{F_{xx}} \Delta \alpha
    \label{eq:S_pheno_short_term_resp}
\end{align}

To derive the slow genetic change after the shift, we expand the fitness gradient along $\overline{y}$ at $t=0^+$:
\begin{align}
    \left. \frac{\partial F}{\partial \overline{y}} \right|_{t=0^+} &= \left. \frac{\partial F}{\partial \overline{y}} \right|_{t=0^-} + \frac{\partial}{\partial \overline{x}} \left( \frac{\partial F}{\partial \overline{y}} \right) \Delta \overline{x} + \frac{\partial}{\partial \overline{y}} \left( \frac{\partial F}{\partial \overline{y}} \right) \Delta \overline{y} + \frac{\partial}{\partial \alpha} \left( \frac{\partial F}{\partial \overline{y}} \right) \Delta \alpha \nonumber \\
    &= F_{xy} \Delta \overline{x} + F_{yy} \Delta \overline{y} + F_{y \alpha} \Delta \alpha
    \label{eq:S_slope_expand}
\end{align}
Under Assumption 3 (quadratic fitness function), higher-order terms are zero, making this first-order expansion exact. In the short-time response, $\Delta \overline{y} = 0$, and the initial change of $\overline{x}$ is solely $\Delta \overline{x}_\mathrm{plas}$. By Assumption 1, the environmental perturbation does not directly create a fitness-gradient component along the slow genotypic coordinate. In the local quadratic expansion, this condition is expressed as $F_{y\alpha}=0$. Thus, immediately after the environmental shift, the genetic gradient is generated by the plastic displacement of the fast phenotypic variable:
\begin{equation}
    \left. \frac{\partial F}{\partial \overline{y}} \right|_{t=0^+} = F_{xy} \Delta \overline{x}_\mathrm{plas}
    \label{eq:S_slope_short_time}
\end{equation}
Substituting this into the evolutionary rate defined in Eq.~(\ref{eq:S_price_evo_Lande}) gives the initial velocity of the evolutionary change:
\begin{equation}
    \left. \frac{d\overline{x}_\mathrm{evo}}{dt} \right|_{t=0^+} = -G_{yy} \frac{F_{xy}^2}{F_{xx}} \Delta \overline{x}_\mathrm{plas}
    \label{eq:S_pheno_initial_velocity}
\end{equation}

\subsection*{Response at an arbitrary time and steady state}
To evaluate the response at an arbitrary time $t$, we integrate the total rate of phenotypic change to decompose the total phenotypic change $\Delta \overline{x}(t) = \overline{x}(t) - \overline{x}(0^-)$ into genetic and plastic displacements:
\begin{align}
    \Delta \overline{x}(t) &= \int_{0^-}^{t} \frac{d\overline{x}_\mathrm{evo}}{dt} dt + \int_{0^-}^{t} \frac{d\overline{x}_\mathrm{plas}}{dt} dt \nonumber \\
    &= \Delta \overline{x}_\mathrm{evo}(t) + \Delta \overline{x}_\mathrm{plas} \nonumber \\
    &= -\frac{F_{xy}}{F_{xx}} \Delta \overline{y} (t) + \Delta \overline{x}_\mathrm{plas}
    \label{eq:S_pheno_delta_arbitrary}
\end{align}
where $\Delta \overline{y}(t) = \overline{y}(t) - \overline{y}(0^-)$. We expand the slope of the fitness landscape using $\Delta \overline{x}(t)$ and $\Delta \overline{y}(t)$:
\begin{equation}
    \frac{\partial F}{\partial \overline{y}} (t) = F_{xy} \Delta \overline{x}(t) + F_{yy} \Delta \overline{y}(t)
    \label{eq:S_slope_decomp_arbitrary}
\end{equation}
Substituting Eq.~(\ref{eq:S_pheno_delta_arbitrary}) into Eq.~(\ref{eq:S_slope_decomp_arbitrary}) gives:
\begin{align}
    \frac{\partial F}{\partial \overline{y}} (t) &= F_{xy} \left( -\frac{F_{xy}}{F_{xx}} \Delta \overline{y} (t) + \Delta \overline{x}_\mathrm{plas}\right) + F_{yy} \Delta \overline{y}(t) \nonumber \\
    &= F_{xy} \Delta \overline{x}_\mathrm{plas} + \left( F_{yy} - \frac{F_{xy}^2}{F_{xx}}\right) \Delta \overline{y} (t) \nonumber \\
    &= F_{xy} \Delta \overline{x}_\mathrm{plas} + \frac{\det H}{F_{xx}} \Delta \overline{y} (t) 
    \label{eq:S_slope_arbitrary}
\end{align}
At the final steady state ($t \rightarrow \infty$), genetic evolution ceases, and $\frac{\partial F}{\partial \overline{y}} (\infty) = 0$. Substituting Eq.~(\ref{eq:S_slope_arbitrary}) into this condition yields the total change in genotype:
\begin{align}
    0 &= F_{xy} \Delta \overline{x}_\mathrm{plas} + \frac{\det H}{F_{xx}} \Delta \overline{y}(\infty) \nonumber \\
    \Delta \overline{y}(\infty) &= - \frac{F_{xx} F_{xy}}{\det H} \Delta \overline{x}_\mathrm{plas}
    \label{eq:S_geno_steady_state}
\end{align}
The total long-time change in the phenotype, $\Delta \overline{x}_\mathrm{total}$, is obtained by evaluating Eq.~(\ref{eq:S_pheno_delta_arbitrary}) at $t \rightarrow \infty$:
\begin{align}
    \Delta \overline{x}_\mathrm{total} &= \Delta \overline{x}_\mathrm{evo}(\infty) + \Delta \overline{x}_\mathrm{plas} \nonumber \\
    &= - \frac{F_{xy}}{F_{xx}} \Delta \overline{y}(\infty) + \Delta \overline{x}_\mathrm{plas} \nonumber \\
    &= \left(1 + \frac{F_{xy}^2}{\det H} \right) \Delta \overline{x}_\mathrm{plas}
    \label{eq:S_le_chatelier_eq}
\end{align}
Because $\det H > 0$, we obtain the inequality $|\Delta \overline{x}_\mathrm{total}| \geq |\Delta \overline{x}_\mathrm{plas}|$.

\section*{S3. Exponential decay and mutation-selection balance}

This section provides the detailed derivations of the exponential relaxation and mutation--selection-balance results presented in the main text.

\subsection*{Derivation of the evolutionary dynamics and relaxation time}
The evolutionary process of the genotype is driven by the Price equation derived in Eq.~(\ref{eq:S_cov_y_expand}). Because $\overline{y} (0^+)$ is constant, we can rewrite $\frac{d \overline{y}}{dt}$ as $\frac{d}{dt} \Delta \overline{y} (t)$. Substituting the slope at an arbitrary time (Eq.~\ref{eq:S_slope_arbitrary}) yields:
\begin{align}
    \frac{d}{dt} \Delta \overline{y} (t) &= G_{yy} \left( F_{xy} \Delta \overline{x}_\mathrm{plas} + \frac{\det H}{F_{xx}} \Delta \overline{y} (t) \right) \nonumber \\
    &= G_{yy} \frac{\det H}{F_{xx}} \Delta \overline{y} (t) + G_{yy} F_{xy} \Delta \overline{x}_\mathrm{plas} \nonumber \\
    &= G_{yy} \frac{\det H}{F_{xx}} \Delta \overline{y} (t) - G_{yy} \frac{\det H}{F_{xx}} \Delta \overline{y}(\infty) \nonumber \\
    &= - \frac{1}{\tau} \left( \Delta \overline{y} (t) - \Delta \overline{y}(\infty) \right)
    \label{eq:S_geno_dynamics}
\end{align}
In the third line, we used the steady-state relationship (Eq.~\ref{eq:S_geno_steady_state}) to eliminate $\Delta \overline{x}_\mathrm{plas}$. In the final line, we defined the relaxation time as $\tau = - \frac{F_{xx}}{G_{yy} \det H}$.

\subsection*{Derivation of the steady-state genetic variance}
We consider the effective fitness landscape experienced by $y$ when $x$ is optimized. By differentiating the expanded slope (Eq.~\ref{eq:S_slope_arbitrary}) with respect to $\overline{y}$, the effective curvature is:
\begin{equation}
    \frac{\partial^2 F}{\partial \overline{y}^2} = \frac{\det H}{F_{xx}} = - \kappa
    \label{eq:S_slope_effective_curvature}
\end{equation}
where $\kappa = -\frac{\det H}{F_{xx}} > 0$. The effective fitness function around the optimal $y_\mathrm{opt} = \overline{y}(0^-) + \Delta \overline{y}(\infty)$ is a quadratic function $F(y) = F(y_\mathrm{opt}) - \frac{\kappa}{2} \left( y- y_\mathrm{opt} \right)^2$. 

Mapping the continuous-time dynamics to discrete generations, the absolute fitness is $\exp(F(y))$. Selection multiplies the Gaussian distribution $P(y)$ (Assumption 4') by a Gaussian weighting factor:
\begin{equation}
    P'(y) \propto P(y) \exp\left[-\frac{\kappa}{2} \left(y- y_\mathrm{opt} \right)^2 \right]
    \label{eq:S_var_dist_update}
\end{equation}
The product of two Gaussian distributions yields a new Gaussian distribution $P'(y)$ with updated variance $G'_{yy}$:
\begin{align}
    \frac{1}{G'_{yy}} &= \frac{1}{G_{yy}} + \kappa \nonumber \\
    G'_{yy} &= \frac{G_{yy}}{1 + \kappa G_{yy}}
    \label{eq:S_var_update}
\end{align}

Under Assumption 5 (mutation-selection balance), $G_{yy} - G_{yy}' = V_m$. Substituting Eq.~(\ref{eq:S_var_update}) yields:
\begin{align}
    G_{yy} &= \frac{G_{yy}}{1 + \kappa G_{yy}} + V_m \nonumber \\
    0 &= \kappa G_{yy}^2 - \kappa V_m G_{yy} - V_m \nonumber \\
    G_{yy} &= \frac{V_m}{2} + \sqrt{\frac{V_{m}^2}{4} + \frac{V_m}{\kappa}}
    \label{eq:S_var_exact_solution}
\end{align}
Assuming the mutational variance is very small ($V_m \ll 1$), we obtain the approximated steady-state variance:
\begin{equation}
    G_{yy} \simeq \sqrt{\frac{V_m}{\kappa}} = \sqrt{-\frac{V_{m} F_{xx}}{\det H}}
    \label{eq:S_var_approx_solution}
\end{equation}

\end{document}